\documentclass{aastex}
\usepackage{natbib}   
\begin{document}
\title{TRIDENT: an Infrared Differential Imaging Camera Optimized for the Detection of Methanated Substellar Companions}

\author{Christian Marois$^{1,2}$, Ren\'{e} Doyon$^{1}$, Daniel Nadeau$^{1}$, Ren\'{e} Racine$^{1}$,\\ Martin Riopel$^{1}$, Philippe Vall\'{e}e$^{1}$, David Lafreni\`{e}re$^{1}$}
\affil{$^{1}$D\'{e}partement de physique, Universit\'{e} de Montr\'{e}al, C.P. 6128, Succ. A,\\ Montr\'{e}al, QC, Canada H3C 3J7 \\$^{2}$ Institute of Geophysics and Planetary Physics, Lawrence Livermore National Laboratory,\\ 7000 East Ave L-413, Livermore CA 94550}
\email{cmarois@igpp.ucllnl.org doyon@astro.umontreal.ca nadeau@astro.umontreal.ca racine@astro.umontreal.ca riopel@astro.umontreal.ca vallee@astro.umontreal.ca david@astro.umontreal.ca}

\begin{abstract}
A near-infrared camera in use at the Canada-France-Hawaii Telescope (CFHT) and at the 1.6-m telescope of the Observatoire du Mont-M\'{e}gantic is described. The camera is based on a Hawaii-1 1024$\times$1024 HgCdTe array detector. Its main feature is to acquire three simultaneous images at three wavelengths across the methane absorption bandhead at 1.6~$\mu$m, enabling, in theory, an accurate subtraction of the stellar point spread function (PSF) and the detection of faint close methanated companions. The instrument has no coronagraph and features fast data acquisition, yielding high observing efficiency on bright stars. The performance of the instrument is described, and it is illustrated by laboratory tests and CFHT observations of the nearby stars GL526, $\upsilon$~And and $\chi$~And. TRIDENT can detect ($6\sigma$) a methanated companion with $\Delta H$~=~9.5 at $0.5^{\prime \prime}$ separation from the star in one hour of observing time. Non-common path aberrations and amplitude modulation differences between the three optical paths are likely to be the limiting factors preventing further PSF attenuation. Instrument rotation and reference star subtraction improve the detection limit by a factor of 2 and 4 respectively. A PSF noise attenuation model is presented to estimate the non-common path wavefront difference effect on PSF subtraction performance.
\end{abstract}

\noindent {\em Keywords}: Instrumentation: adaptive optics - infrared: stars - planetary systems - stars: low-mass, brown dwarfs.\\

\section{Introduction}

The search for substellar companions (brown dwarfs and exoplanets) around nearby stars is essential to better understand their formation history, physical characteristics and to establish whether our Solar system is typical. Following the discovery of the first confirmed brown dwarf \citep{nakajima1995}, several surveys \citep{burgasser2000,oppenheimer2001,luhman2002,metchev2002,patience2002,kaisler2003,mccarthy2004} have searched the Solar neighborhood and found several brown dwarf companions. Detection at small separations of fainter, Jupiter mass, companions having a planet to star brightness ratio as low as 10$^{-8}$ is more difficult. Current instruments used with 4 to 10-m telescopes and adaptive optics (AO) systems can detect, at 0.5$^{\prime \prime}$ separation, companions with a brightness ratio of $\sim $~10$^{-4}$. The main limitation is the structure of the stellar point spread function (PSF) that masks the faint substellar companion signal.

PSF subtraction is a challenging endeavour. Atmospheric speckles \citep{racine1999} and quasi-static instrumentally induced aberrations \citep{marois2003a} produce a PSF that changes with time, limiting the efficiency of PSF subtraction. Simultaneous spectral differential imaging (SSDI) is a promising technique to calibrate both of these effects and achieve photon noise limited detections \citep{smith1987,racine1999,marois2000a,sparks2002,close2005}. It consists of simultaneously acquiring images at adjacent wavelengths through a spectral range where the companion and stellar spectra differ appreciably, and combining the images in a way that separates the stellar and companion signals \citep{marois2000a}. A good spectral feature for SSDI is the sharp methane absorption bandhead at 1.6~$\mu$m \citep{rosenthal1996} found only in relatively cold atmospheres ($<$1300~K) \citep{burgasser2002} such as those of T type brown dwarfs and Jovian planets.

This paper describes the TRIDENT near-infrared camera based on the SSDI concept presented in \citet{marois2000a,marois2000b}. Details about the camera design, laboratory tests, results on CFHT along with performance simulations are presented.

\section{Description of the Camera}

\begin{figure}
\plotone{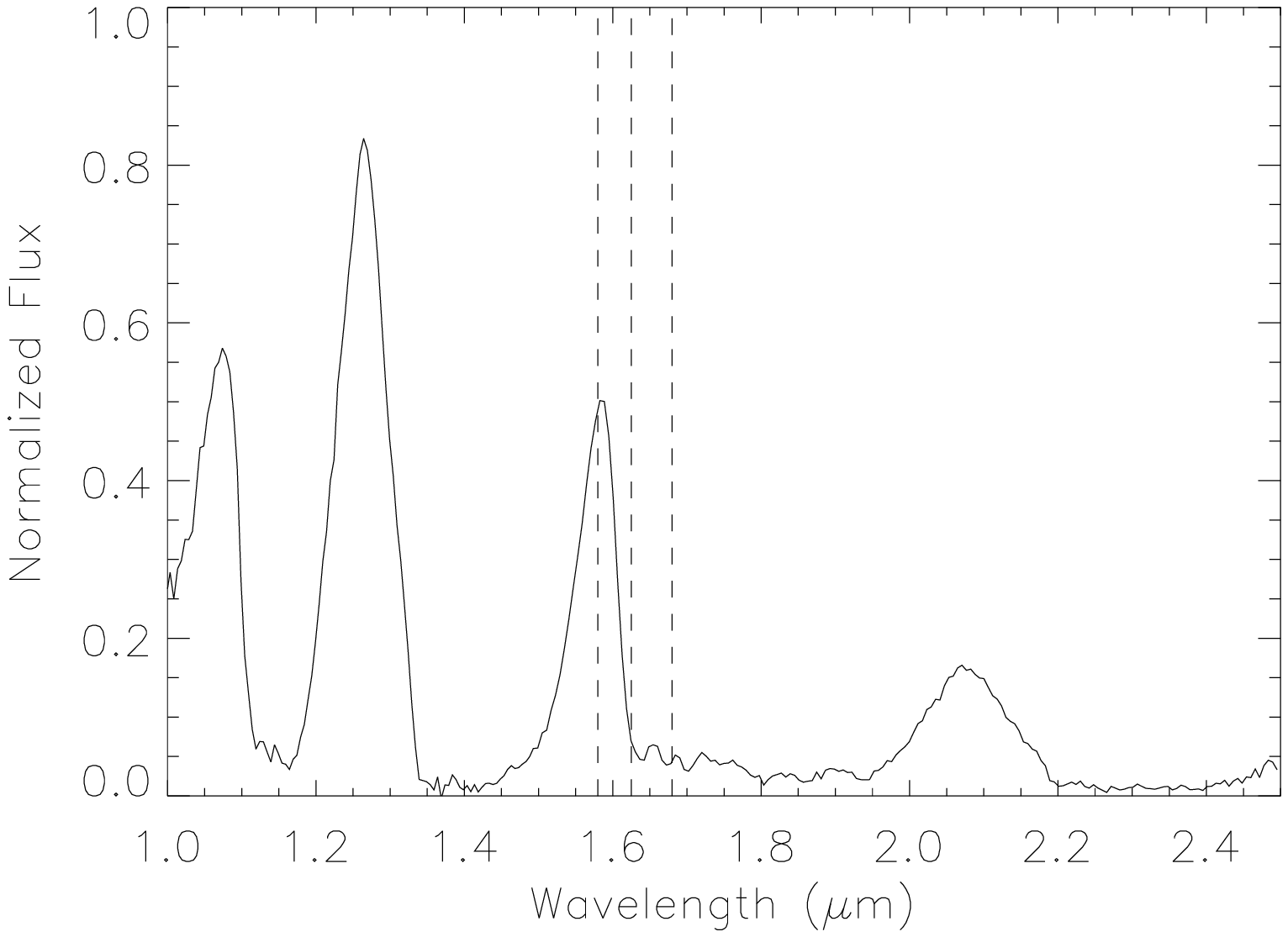}
\caption{Near-IR spectrum of the T8 brown dwarf 2MASS~0415-0935 \citep{burgasser2002}. The three dashed lines represent the central wavelength of the three TRIDENT filters (1.580~$\mu$m, 1.625~$\mu$m and 1.680~$\mu$m, 1\% bandwidth).\label{pasp2003fig0}}
\end{figure}

TRIDENT is designed to operate with an adaptive optics (AO) system (PUEO at CFHT, \citep{Rigaut98} or the OMM AO system \citep{ivanescu2003}) to benefit from diffraction-limited images. The goal was to build a compact and simple instrument that would be easy to carry from the laboratory to any observatory. Commercial components were used as much as possible to minimize development time. The camera features a special polarizing beam-splitter allowing the acquisition of simultaneous images in three distinct narrow spectral bands. The wavelengths (1.580~$\mu$m, 1.625~$\mu$m and 1.680~$\mu$m, 1\% bandwidth) have been selected across the 1.6~$\mu$m methane absorption bandhead (see~Fig.~\ref{pasp2003fig0}). The filter centered at 1.580~$\mu $m minimizes the stellar-to-methanated companion flux ratio.

\subsection{Optical Design}
\begin{figure}
\epsscale{0.5}
\plotone{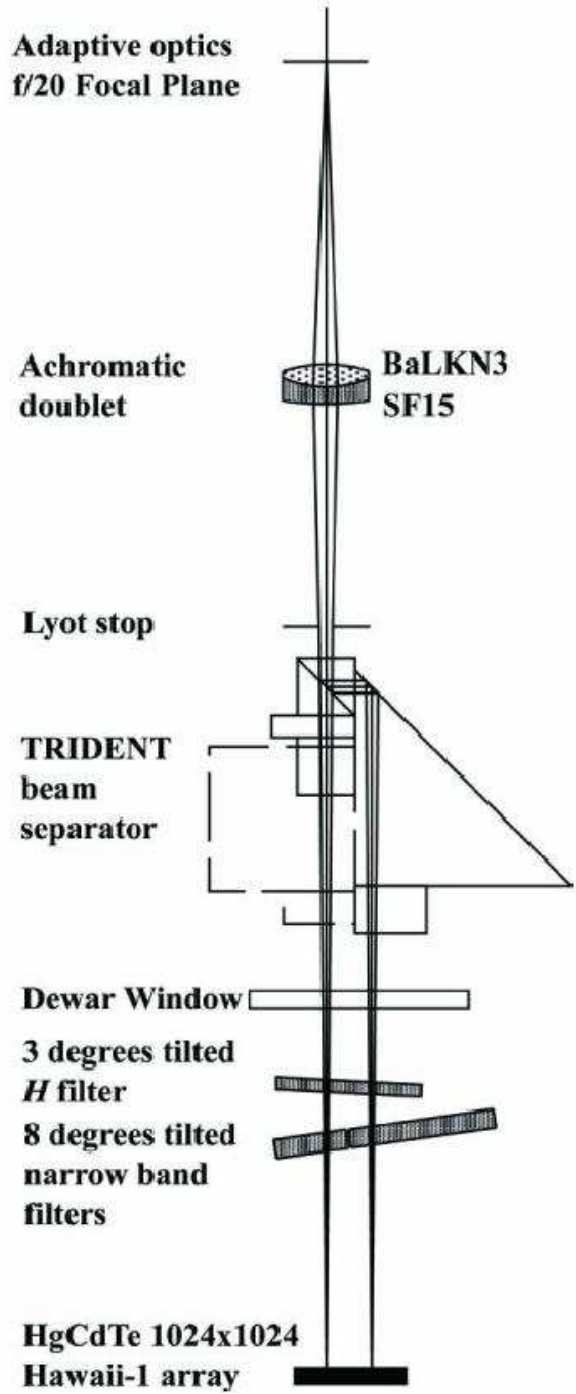}
\caption{Optical layout of TRIDENT. The distance from the adaptive optics focal plane to the detector surface is 228~mm. The optical paths for two of the three simultaneous images are seen in the representation. The third one is in a perpendicular view not visible in this figure. See text for a detailed description.\label{pasp2003fig2}}
\end{figure}

An optical layout of the instrument is shown in Fig.~\ref{pasp2003fig2}. The optical train is in two parts: i) a field stop, achromatic doublet lens (Melles Griot LAL11), Lyot stop, beam separator and BK7 cryostat window, all at room temperature, and ii) an $H$ band filter to block radiation longward of 1.8~$\mu $m, three narrow-band (1\%) filters, and the detector, inside the cryostat. The diameter of the circular field stop is chosen to obtain the maximum field of view (FOV) with minimum overlap. The achromatic doublet re-images the AO focal plane on the detector and creates an image of the pupil on the 2~mm diameter Lyot stop. The designed image scale is $0.018^{\prime \prime}$ pixel$^{-1}$ at CFHT and $0.038^{\prime \prime}$ pixel$^{-1}$ at OMM, corresponding to $\sim$~5 pixels/FWHM. The main reason for such an oversampling is to ensure accurate PSF rescaling and registering and avoid potential interpolation noise that could mask faint companions.

The beam separator uses a combination of two polarizing beam splitters, two right angle prisms and a first order quarter wave retarder to generate three optical beams, organized in a ``L'' shape, each forming an image on one quadrant of the detector, so that all three can be read simultaneously. The first beam is reflected off a beam splitter and a prism to the entrance of the cryostat. The transmitted light has its polarization changed from linear to circular by the retarder. The second beam then goes through the second beam splitter and to the entrance of the cryostat. The third beam is reflected off this beam splitter and a prism to the entrance of the detector. Thin sheets of glass were cemented under the two prisms to ensure co-focality of the three channels. All optical components of the beam separator were optically cemented together using a Norland optical Adhesive with UV curing.

Light enters the cryostat and goes through an $H$ band and narrow-band filters to the detector. All filters were manufactured by Barr Associates. The $H$ band filter is tilted by 3 degrees and the narrow-band filters by 8 degrees to make most of the ghosts fall outside the FOV. The brightest remaining ghost has an intensity of 0.8\% of the actual PSF and it is located $0.46^{\prime \prime}$ from the PSF core. The ghosts have a different intensity and location in each optical channel.  They will thus leave residuals when combining images, but being faint and small ($\sim \lambda$/D), they only affect a small portion of the FOV at known positions.

\subsection{Mechanical Design}

\begin{figure}
\epsscale{1}
\plotone{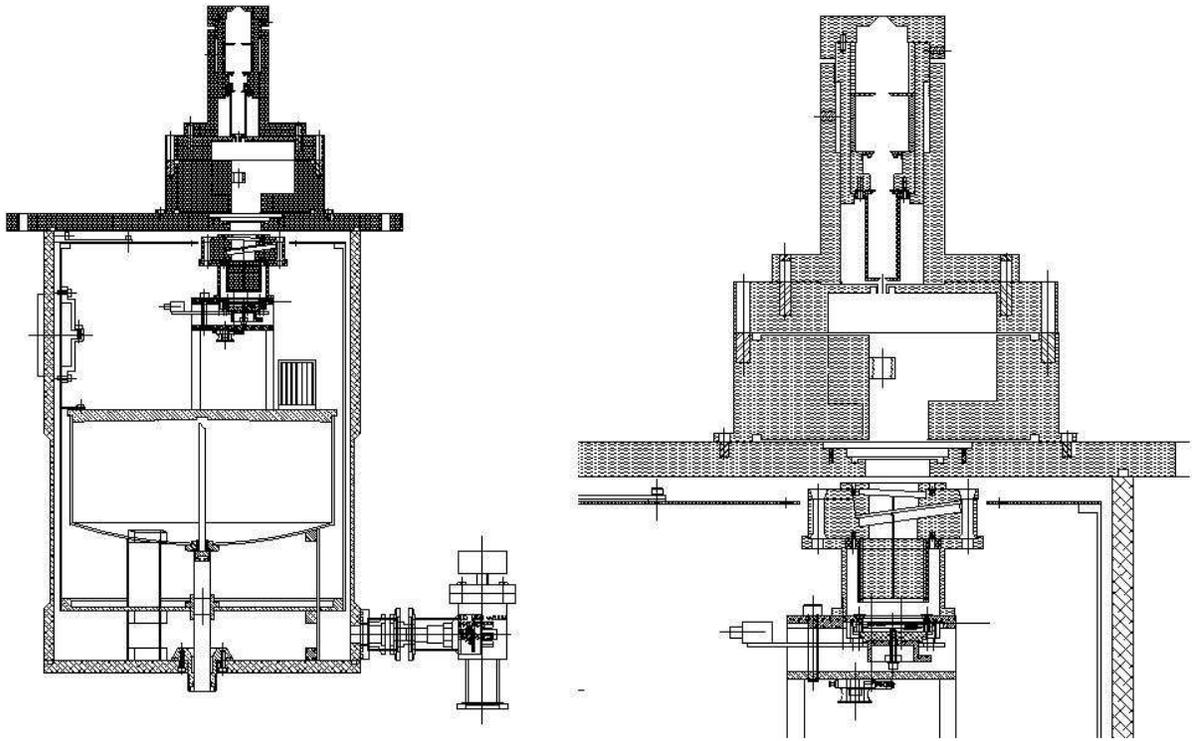}
\caption{Cross section of the TRIDENT camera. Left: global view of the instrument. Right: details of the opto-mechanical bench.\label{pasp2003fig3}}
\end{figure}

A mechanical layout of the cryostat is shown in Fig.~\ref{pasp2003fig3}. The camera is housed in an Infrared Laboratories\footnote{Tucson, AZ, http://www.irlabs.com/} ND8 cryostat having a diameter of 21~cm and an overall height of 32~cm. It contains a single liquid nitrogen reservoir with a hold time of 50~h. A temperature controller (Lakeshore model 331) is used to stabilize the detector temperature at $80$~K within mK accuracy. There is no moving part.

\subsection{Data Acquisition System and Software}

The detector is a Hawaii-1 1024$\times$1024 HgCdTe array \citep{hodapp1996}, mounted on a IRLab fanout board. The four channels, one for each quadrant, are first pre-amplified in parallel just outside the cryostat and then fed to an SDSU-2 controller \citep{leach1996} where the analog-to-digital conversion takes place. The measured read noise is 18~e$^{-}$. The minimum conversion time of 1~$\mu$s is used for reading the array as fast as possible to allow unsaturated observations of bright stars with high efficiency. The frame read time is 0.262~s. On board image co-addition enables Fowler sampling \citep{Fowler1990} to minimize readout noise. The system also features a unique clocking pattern that eliminates the reset anomaly of the Hawaii array without loss of observing time \citep{Riopel2004}. These features make the data-acquisition process very efficient and simple.

The host computer is a SUN Microsystem Ultra~5 with a PCI acquisition board connected to the SDSU-2 via a 40~Mbits/sec fiber optic link for fast data transfer (0.75~s per 1024$\times$1024 pixel, 32~bit image).

\begin{figure}
\plotone{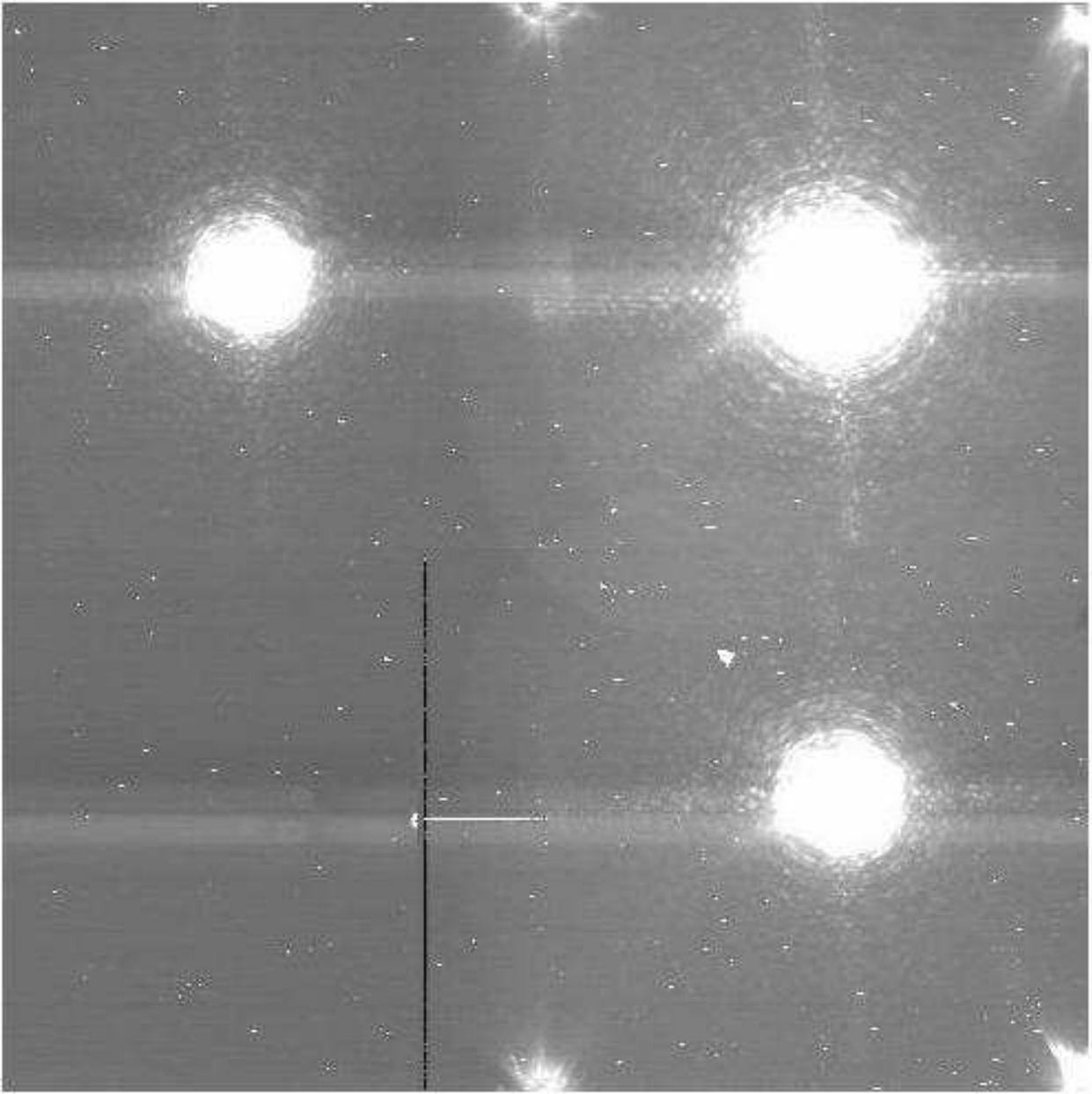}
\caption{A typical unprocessed TRIDENT image obtained at CFHT with PUEO. The upper left, upper right, and lower right quadrants display the 1.580~$\mu$m, the 1.680~$\mu$m, and the 1.625~$\mu$m images respectively. Crosstalk (horizontal bright features) is visible across quadrants. The FOV is $9^{\prime \prime}$ per quadrant at CFHT and $18^{\prime \prime}$ at OMM. Amplifier glow is seen in either the lower right or upper right corner of each quadrant.\label{pasp2003fig4}}
\end{figure}

During an observing session, control is done from a single computer display. A C-based program with Tcl/Tk user interface with a shared memory link to the SAOimage DS9 display program controls image acquisition and display. The data are stored on disk as 32 bit integers in FITS format \citep{Wells1981,hanisch2001} with a header that includes all the observational parameters. The computer can be used at any time to look at recently acquired images to ascertain the data quality with the DS9 display program. An example of a TRIDENT image is shown in Fig.~\ref{pasp2003fig4}. Images can be downloaded to another computer in the local network for further image analysis. An IDL-based software reduction pipeline is used to do image reduction and analysis (dark subtraction, flat field normalization, bad pixel correction and image trimming, registration, scaling and subtraction) while observing at the telescope. 

\section{Instrument Performance}

The rationale behind the design of TRIDENT is to make possible the coherent subtraction of atmosphere induced speckles. The timescale of variability is so short that simultaneous measurements through different optical channels is the only practical solution. The introduction of differential aberrations of the wavefront 
between channels is thus unavoidable, its extent depending on both the design and its implementation. Laboratory tests to assess the extent of such differential aberrations in TRIDENT are presented in this section. The second step requires observations at the telescope to determine how well and by what method such aberrations can be calibrated and subtracted at the same time as atmosphere induced aberrations. Such observations are presented in the second part of this section.

\subsection{Laboratory Performance\label{seclabo}}
TRIDENT was tested in the laboratory with its 2~mm diameter Lyot stop in place and a 3~$\mu$m pinhole situated at the entrance focal plane to produce a 5 pixels/FWHM PSF. Each sequence of data was the result of $30\times20$~sec exposures, the integration time being chosen so as to remain well within the linear regime of the detector.  In total, 10 such sequences were acquired to get sufficient signal to noise inside a 5 $\lambda$/D radius ($0.5''$ radius at CFHT), the region of greatest interest for the faint companion search.

Dark frames having the same exposure time and number of multiple sampling readouts were subtracted from images and flat fields. Each image was then divided by the combined flat field image. Bad and hot pixels were corrected by interpolating from nearby pixels. The 512$\times$512 pixel images at each of the three wavelengths were then extracted from the original 1024~$\times$~1024 pixel images. A Fourier space algorithm was used to accurately register all three PSFs to a common center \citep{marois2004c}.

Strehl ratios for the three optical channels were estimated by comparing the instrumental PSF peak intensities with those of theoretical PSFs convolved by the pinhole diameter and having the same flux normalization and number of pixels per FWHM. The Strehl ratios were found to be $0.95$ with estimated errors of $\pm$~0.03, or $40 \pm 12$~nm rms for 1.6~$\mu $m PSFs.

An empirical fit to the PSF structure can be done using numerical simulations. Assuming Fraunhofer diffraction, the PSF intensity $I$ in image plane coordinates $\eta$ and $\xi$ is simply the Fourier transform of the complex pupil 

\begin{equation}
I(\eta,\xi) = \left| FT \left( A(x,y) e^{i\left[ \phi(x,y)\right]}\right) \right|^2
\end{equation}

\noindent where $A$ and $\phi$ are respectively the pupil amplitude and phase error functions expressed in the pupil plane coordinates $x$ and $y$. Considering a uniform circular pupil, the PSF structure will be generated by phase errors only. Assuming that the spectrum of the phase errors is described by a sum of power laws ($P_i \propto \nu^{\alpha_i}$), a good analytical approximation for polished optical surfaces \citep{church1988}, the phase error function is simply

\begin{equation}
\phi(x,y) = \sum_i \Re \left[ FT \{ P_i(\kappa,\zeta) \}\right]
\end{equation}

\noindent where $P_i$ is a power-law expressed in coordinates $\kappa$ and $\zeta$, the spatial phase error frequencies in cartesian coordinates ($\nu^2 = \kappa^2+\zeta^2$) expressed in units of cycles per pupil diameter. The real part of the Fourier transform is chosen, to produce a real phase error function. The phase error $\phi$ is then normalized to the desired amplitude inside the pupil. Simulations were executed with different combinations of $\alpha$ exponents and phase error amplitudes until an empirical fit was found. A combination of two phase error power-law spectra having $\alpha_1$~=~$-1.0$ and $\alpha_2$~=~$-2.7$ with 45~nm and 20~nm rms respectively, for a combined value of 50~nm rms ($S = 0.96$) closely matches the PSF structure at all radii. This PSF model is in good agreement with the measured Strehl ratio of all three TRIDENT PSFs.

Images at 1.625 and 1.680~$\mu$m were then scaled spatially to 1.580~$\mu$m to account for the PSF chromaticity. Scaling factors obtained empirically were found to be very close to the ratio of observing wavelengths, as one would expect from diffraction theory. Image scaling was achieved using an iterative FFT padding technique \citep{marois2004c}. The three simultaneous images ($I_{\lambda_1}$, $I_{\lambda_2}$ and $I_{\lambda_3}$) are then normalized to the same integrated intensity.

The three images ($I_{\lambda_1}$, $I_{\lambda_2}$ and $I_{\lambda_3}$) are combined following \citet{marois2000a}. First, two simple image differences ($SD_{3,1}$ and $SD_{2,1}$) are evaluated:

\begin{equation}
SD_{j,i} = I_{\lambda_{j}} - I_{\lambda_{i}}\label{pasp2003equ2}
\end{equation}

\noindent A double difference image ($DD$) is then obtained by subtracting the two simple differences:

\begin{equation}
DD = k_2 \left( k_1 SD_{3,1} - SD_{2,1} \right)\label{pasp2003equ3}
\end{equation}

\noindent where $k_1$ and $k_2$ are normalizing constants. The factor $k_1$ normalizes the wavelength differences between the wavelengths of the two SDs. For TRIDENT wavelengths, this factor is 0.45. Since a methanated companion has been chosen to be brigther in the $I_{\lambda_1}$ image, the $DD$ image must be normalized so that methanated companions are as bright in the $DD$ image as in the $I_{\lambda_1}$ image, taken as a reference in the noise assessment below. This additional normalizing factor $k_2$ is found to be $\sim$2. Before a noise comparison is made, an azimuthally averaged profile is subtracted from the $I_{\lambda_1}$, the $SD_{2,1}$, and the $DD$ images. As long as the circumference over which this average is made is significantly greater than the PSF FWHM, this subtraction will have a negligible effect on the amplitude of the signal from a companion.

\begin{figure}
\plotone{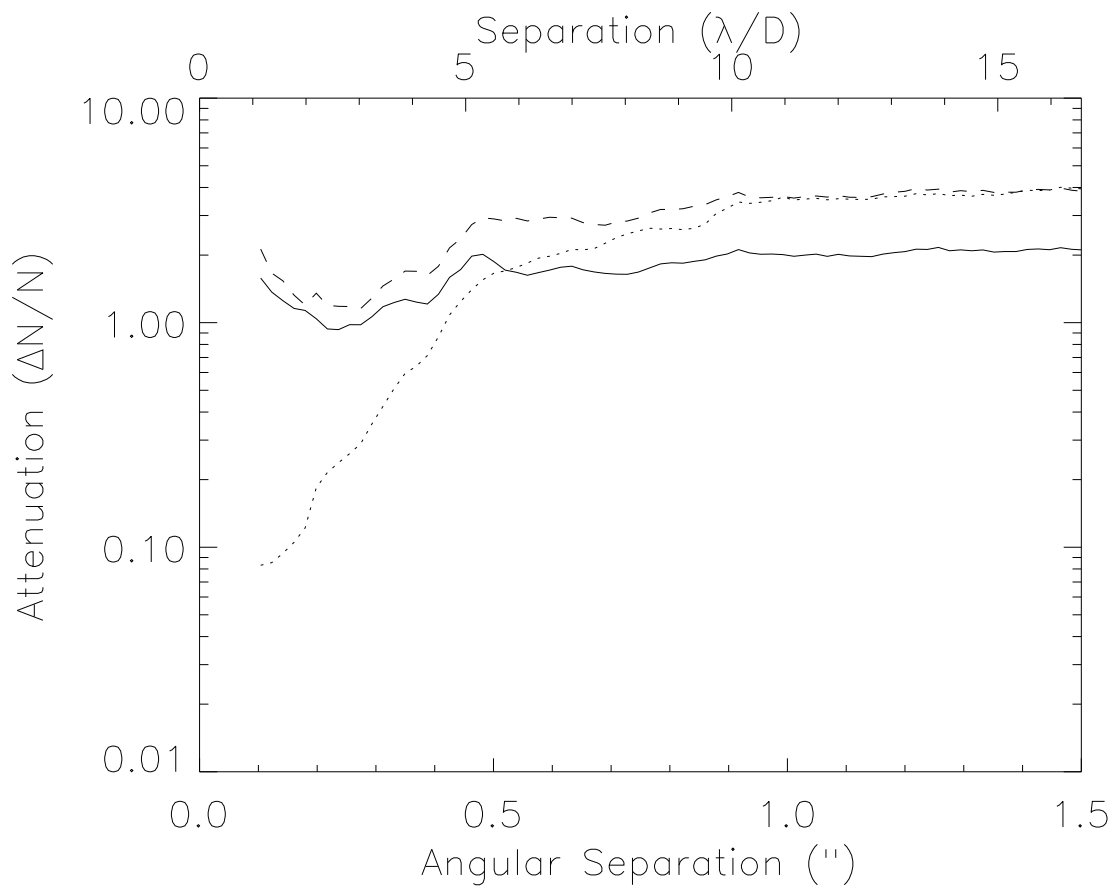}
\caption{Median of the pixel-to-pixel ratio of residual noise to initial noise as a function of the distance from the PSF center. The solid line shows the results from the SD between $I_{\lambda_1}$ and $I_{\lambda_2}$ while the dashed line shows the results from the DD between the $I_{\lambda_1}$, $I_{\lambda_2}$ and $I_{\lambda_3}$ images. The dotted line shows the limit imposed by flat field, photon and read noises for the DD. Attenuation is calculated relative to the noise $N$ of the $I_{\lambda_1}$ image. The horizontal axis scale corresponds to the CFHT data.\label{pasp2003fig5}}
\end{figure}

The level of noise attenuation is examined as a function of angular separation, defined as the mean radius of an annulus of width equal to the PSF FWHM.  The noise attenuation factor $\Delta N/N$ is defined as the median of the pixel-to-pixel ratio of the absolute value of the signal in a $SD_{j,i}$ or $DD$ image to its absolute value in the $I_{\lambda_1}$ image. Figure~\ref{pasp2003fig5} shows the $SD_{2,1}$ and $DD$ PSF noise attenuations measured in the laboratory. 

For separations greater than $0.6''$, the PSFs are read noise limited. As expected, because of the normalization explained above, in the limit of high angular separation, the $SD_{2,1}$ noise is twice that in $I_{\lambda_1}$ and the $DD$ noise is another factor of 2 above that in the SD. 

For smaller separations, the SD and DD PSF residuals have amplitudes of the same order or larger than the original PSF noise, the instrumental PSF noise attenuation of $SD_{2,1}$ at $0.3''$ being a factor of $\sim3$ from the photon and read noise limit.

\begin{figure}
\plotone{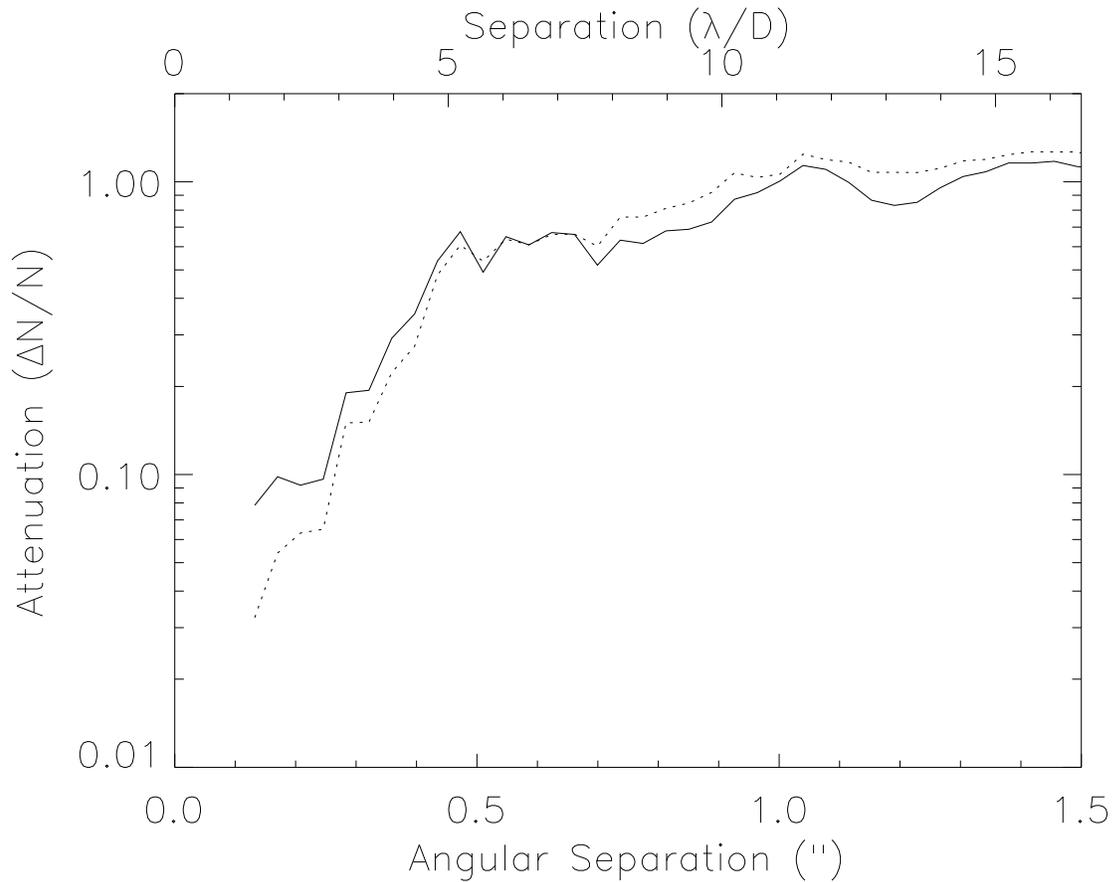}
\caption{Flat field accuracy for TRIDENT laboratory PSFs. One image at one wavelength is subdivided in four separate half-sampled images. The solid line shows the subtraction performance of two such images. The dotted line represents the calculated photon and read noises. The noise at large separation is not $\sqrt{2}$ times the noise of a single image due to partial subtraction of low-frequency detector readout noise.\label{pasp2003fig6}}
\end{figure}

Flat field accuracy is estimated from the laboratory PSF in half-sampled images obtained by selecting one pixel from each 2~$\times$~2 pixel cluster of the original image. Two of those images are registered and subtracted from each other, and the attenuation achieved is shown in Fig.~\ref{pasp2003fig6}. The residual noise after subtraction confirms that flat field noise is not the limiting factor for the noise attenuation.

\citet{perrin2003} have shown that phase errors for unobstructed/unapodized wavefronts (relevant to our laboratory PSF) produce an anti-centrosymmetric PSF structure within a radius of a few $\lambda/D$ when the Strehl ratio is greater than 0.9, while amplitude modulations produce a centrosymmetric PSF structure. Since neither centrosymmetry nor anti-centrosymmetry is seen in the laboratory PSFs, phase errors and amplitude modulations seem to contribute approximately equally to the PSF structure.

These measurements indicate that non-common-path aberrations dominate over common-path aberrations in the laboratory tests, an expected result considering that the beams are separated early in the optical path from the focal plane pinhole to the detector.

\subsection{On-Sky Performance}
Observations were conducted on 2001 July 8-12 and 2001 November 21-24 at the $f/20$ focus of the adaptive optics bonnette PUEO on CFHT \citep{Rigaut98}. Data were obtained on a total of 35~stars with spectral type ranging from B to M.

In July, data well inside the linear regime of the detector were acquired to test the PSF stability.  Observation sequences consisted of exposures lasting a few seconds, coadded to produce an image every minute; this was repeated to reach a total integration time of 30 minutes to 1 hour per target.  Seeing conditions for this run were average to good (Strehl of 0.2-0.5 in the $H$ band). In November, typically, three 1 minute long coadditions of unsaturated exposures lasting a few seconds each were followed by 15 1 minute long coadditions of approximately 15 second exposures, saturated in the core of the primary star.  Total integration time was $\sim 1$~h per target with good seeing conditions (Strehl of 0.5 in $H$). Plate scale and field orientation were determined by observation of the double stars HIP95593 and HIP96570 \citep{perryman1997} in July, while in November, the \citet{mccaughrean1994} TCC040 Orion field was observed. Both measurements yield a plate scale of $0.0189\pm0.0001^{\prime \prime}$/pixel, very close to the expected value.

Dark subtraction, flatfielding, bad pixel correction and image registration were performed as described in the laboratory performance section. Strehl ratios were calculated by normalizing the PSF integrated flux to unity, and dividing its central pixel by that of a theoretical PSF taking into account the telescope central obscuration. If the image is saturated, Strehl ratios are taken to be the average of the previous three non-saturated images. A combined image (weighted by the square of the Strehl ratio) for each wavelength of an object was generated by coadding all images for that object.

The PSF structure can be well approximated using a single power-law phase error distribution with an $\alpha$ exponent of $-2.7$ and 130~nm rms wavefront error. This level of aberration is consistent with the one found by \citet{Rigaut98} from the PUEO performance analysis. Images at 1.625 and 1.680~$\mu$m were then spatially scaled to the 1.580~$\mu$m image using an iterative FFT padding technique \citep{marois2004c}. Finally, the simultaneous images $I_{\lambda_1}$, $I_{\lambda_2}$ and $I_{\lambda_3}$ were normalized and combined following Eq.~\ref{pasp2003equ2} and Eq.~\ref{pasp2003equ3}.

\begin{figure}
\plotone{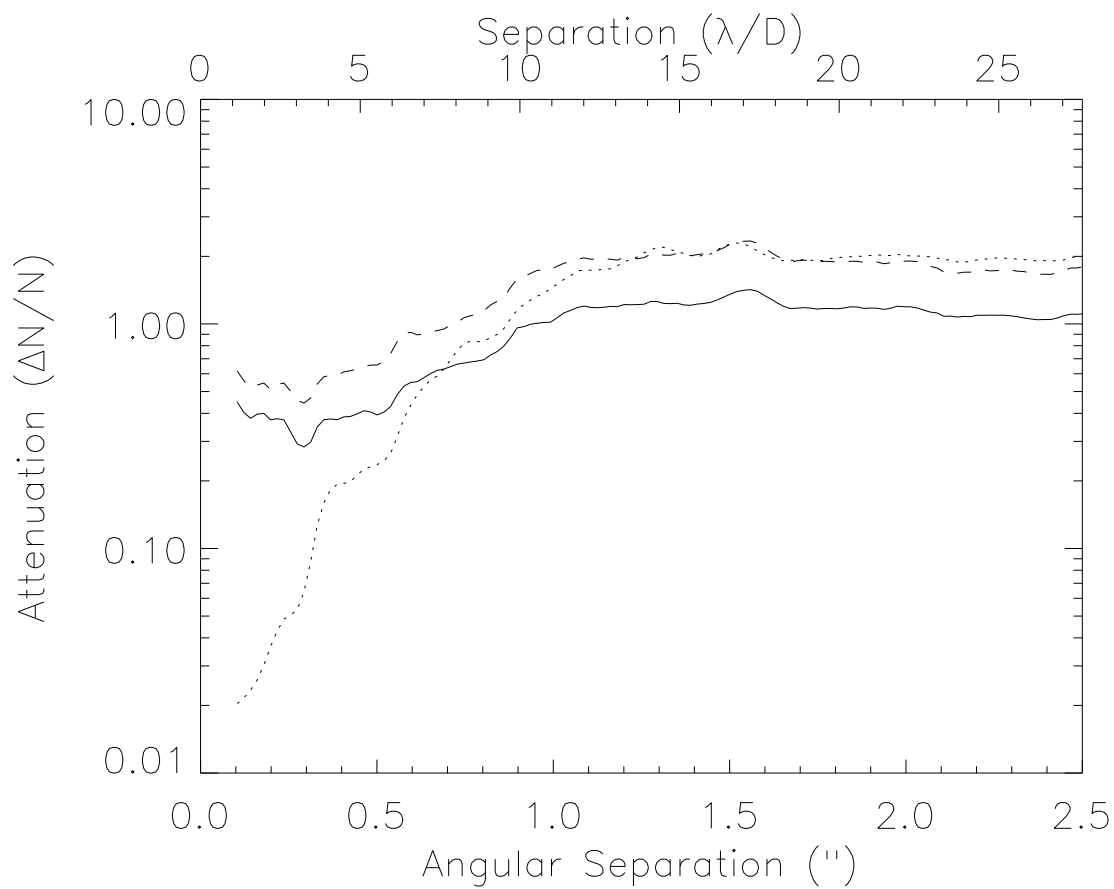}
\caption{PSF noise attenuation with separation obtained with the TRIDENT camera at CFHT for the star GL526. The solid line represents the SD between $I_{\lambda_1}$ and $I_{\lambda_2}$ while the dashed line is for the DD. The dotted line represents the flat field accuracy, photon and read noise limit for the DD.\label{pasp2003graphgl526}}
\end{figure}

The performance is illustrated by data obtained in 1~hour of integration time during the night of 2001 July~8, on the star GL526. Fig.~\ref{pasp2003graphgl526} shows the PSF noise attenuation with separation for both the $SD_{2,1}$ and $DD$ images, compared with the estimated sum of photon and read noise for the $DD$ image. The $SD_{2,1}$ noise is attenuated by a factor of 2.5 inside a $0.5''$ radius, down to 1.0 at a $1.0''$ radius.  The $DD$ image is noisier than the $SD_{2,1}$ image to the same degree as for laboratory measurements, and only the $SD_{2,1}$ image is considered in the subsequent analysis.

\begin{figure}
\plotone{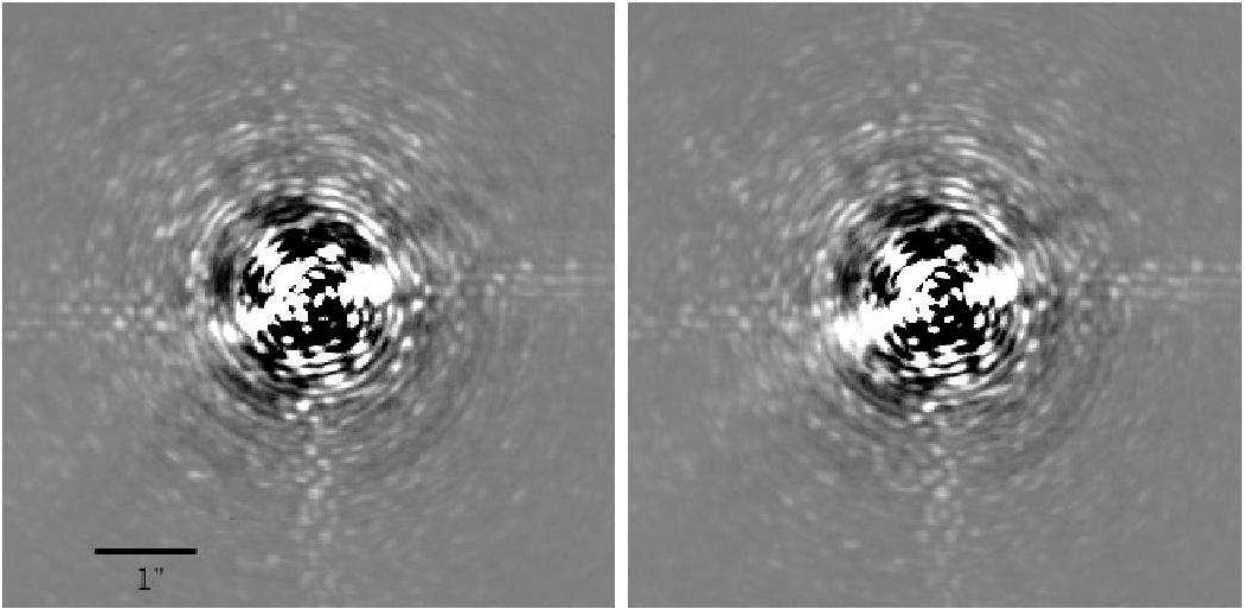}
\caption{Two 20 minute integrations, separated by a 50 minute interval, acquired in the same bandpass during the night of 2001 November 21, on the star $\upsilon$~And. Azimuthally averaged radial profiles have been subtracted. The data are shown on an intensity scale linear within a range of $\pm3\times10^{-4}$ times the star peak intensity; the field of view shown is $5.5^{\prime \prime} \times 5.5^{\prime \prime}$. \label{pasp2003fig8struc}}
\end{figure}

On short timescales, the PSF structure varies with the atmospheric speckles, but it converges to a specific pattern after a few minutes of integration (see Fig.~\ref{pasp2003fig8struc}), as it becomes dominated by the more constant telescope and instrument induced aberrations.  Subtraction of a reference PSF acquired simultaneously at another wavelength through a different optical channel attenuates this structure by a factor of $\sim2$ as it removes the common-path contribution to the aberrations. 

\begin{figure}
\plotone{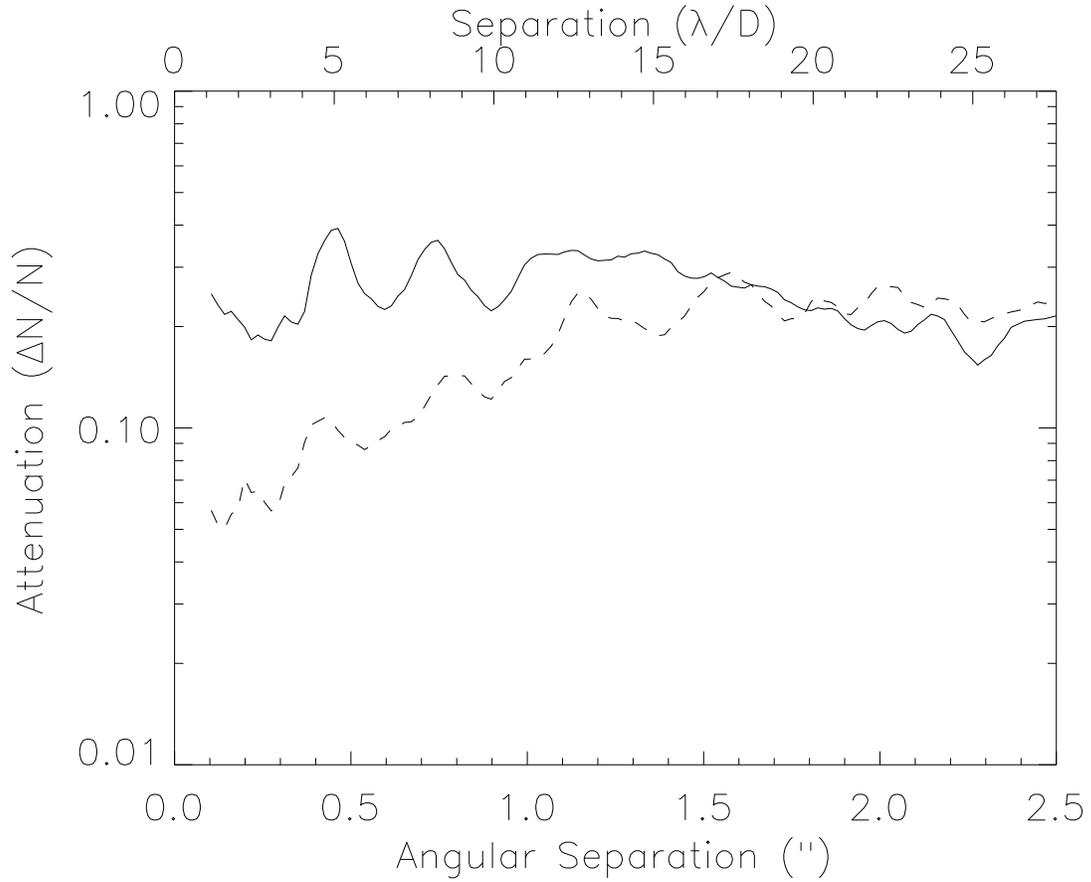}
\caption{The SD PSF noise attenuation with separation is shown for the star $\upsilon$~And for the instrument rotation technique (solid line) after 1h30 of integration time and for the reference star subtraction technique (dashed line, reference star is $\chi$~And) after 1h of integration.\label{pasp2003fig7b}}
\end{figure}

Instrument rotation was performed in 2001 November in order to determine the origin of the instrument induced aberrations. By rotating PUEO and TRIDENT with respect to the telescope, aberrations due to the former stay aligned with the detector, while aberrations from the telescope rotate.  A set of 90~images acquired by steps of 2~degrees from $-90^\circ$ to $+90^\circ$ was obtained. Most of the 130~nm rms instrumental aberrations are coming from PUEO and TRIDENT as the PSF structure is seen to remain fixed with respect to these instruments.  This PSF structure is smoothed when the images are rotated back to align the fields before coaddition, leading to a reduction of the noise by a factor of $\sim2$ at $0.5''$ (see Fig.~\ref{pasp2003fig7b}).

Also in 2001 November, observations of a reference star were made with the aim of calibrating the differential aberrations between the optical channels.  The reference star is selected to be approximately of the same magnitude, spectral type, and declination as the target but offset from it by 10 minutes East or West. This setup ensures that when both objects are acquired 10 minutes apart, the telescope orientation is approximately the same and the PSF remains as stable as possible. As shown in Fig.~\ref{pasp2003fig7b}, reference star subtraction improved the noise attenuation by a factor of $\sim4$ at a $0.3''$ radius.

\begin{figure}
\plotone{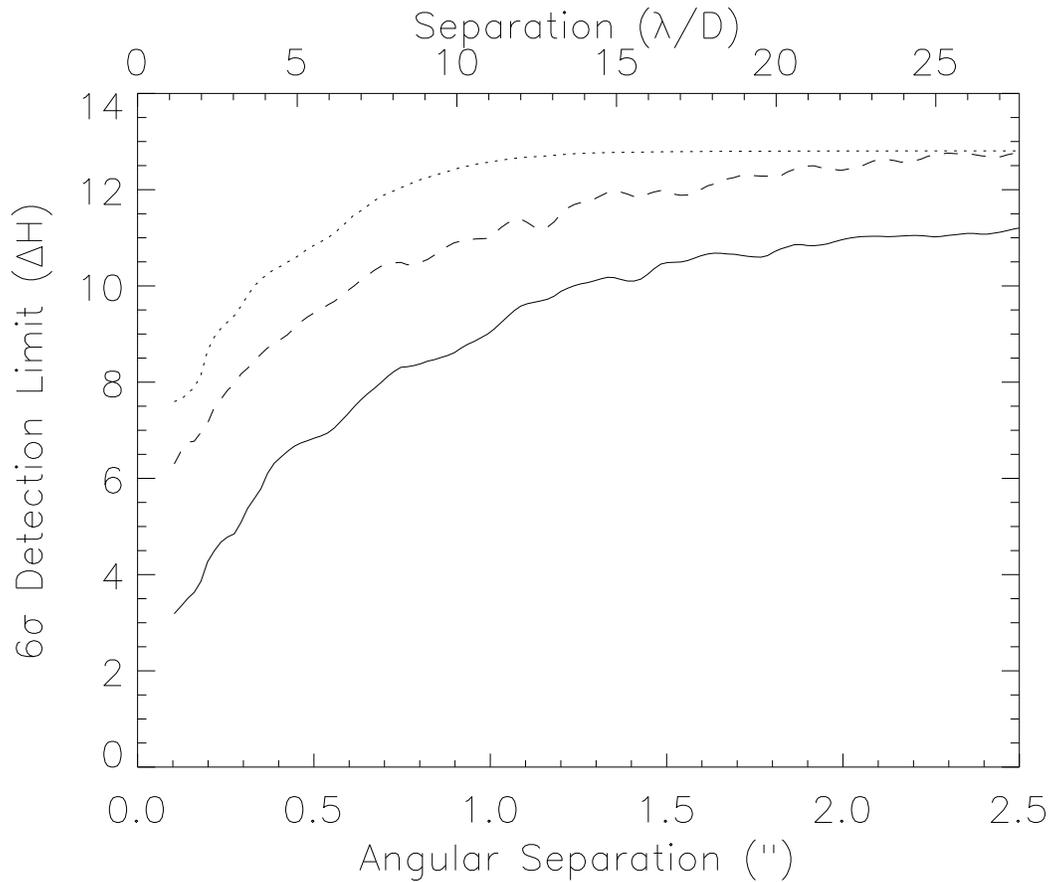}
\caption{Magnitude difference between a star and its companion such that the flux of the companion is $\sim 6$~times the residuals at the corresponding angular separation from the star. The magnitude difference is calculated inside annuli of increasing radius and width $\lambda/D$. The solid line shows the magnitude difference obtained by 2-wavelength subtraction with TRIDENT at CFHT. The dashed line shows the magnitude difference after subtraction of a reference star. The dotted line shows the magnitude difference for data limited by photon and read noise.\label{pasp2003fig7cont}}
\end{figure}

The detection limit (expressed in difference of magnitude) for the reference star subtraction technique is presented in Fig.~\ref{pasp2003fig7cont}. The two targets are $\upsilon$~And and $\chi$~And, acquired during the night of November 21, 2001. The magnitude difference in the $H$ band is shown to facilitate comparison with brown dwarf cooling models \citep{baraffe2003}. With a 1\% bandpass filter, the flux ratio between the star and methanated companion is decreased by a factor $\sim$~3 compared to a broad-band $H$ filter, as inferred from the T8 brown dwarf spectrum of \citet{burgasser2002}. A companion being $\sim $9.5 magnitudes fainter than its star in the $H$ band is detectable at 0.5$^{\prime \prime}$.

\begin{figure}
\plotone{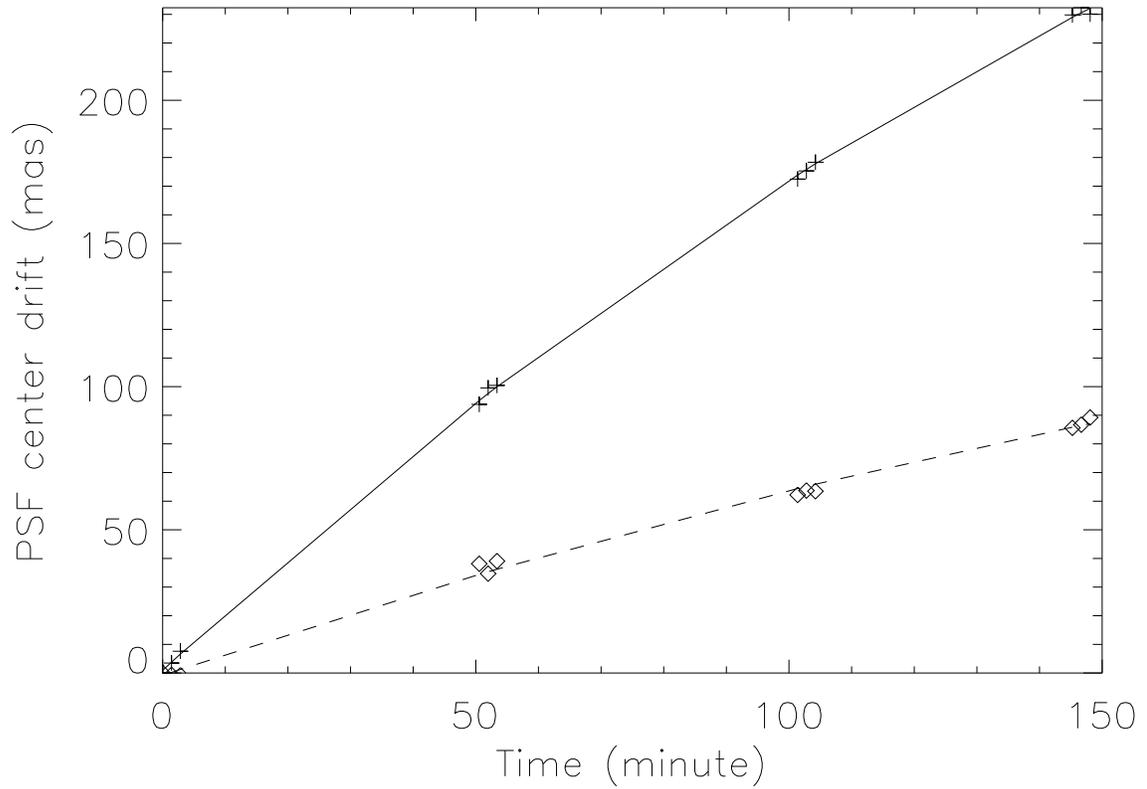}
\caption{Unsaturated PSF center drift for a 2.5~hour integration on $\upsilon$~And acquired during the night of 2001 November 21. The hour angle at the start of the sequence is $-2.5$~h. The solid line is the PSF center drift for a single PSF while the dashed line is the PSF center drift differences between the 1.580~$\mu $m and the 1.680~$\mu $m PSFs multiplied by a factor of 10.\label{pasp2003fig9}}
\end{figure}

Even with proper care in conducting the observations, as described above, the reference PSF subtraction technique appears to be affected by a slow evolution of the differential aberrations. This may be due to relative drifts of the visible AO and infrared wavefronts across instrument optics because of flexures and/or atmospheric refraction (see Fig.~\ref{pasp2003fig9}).

\begin{figure}
\plotone{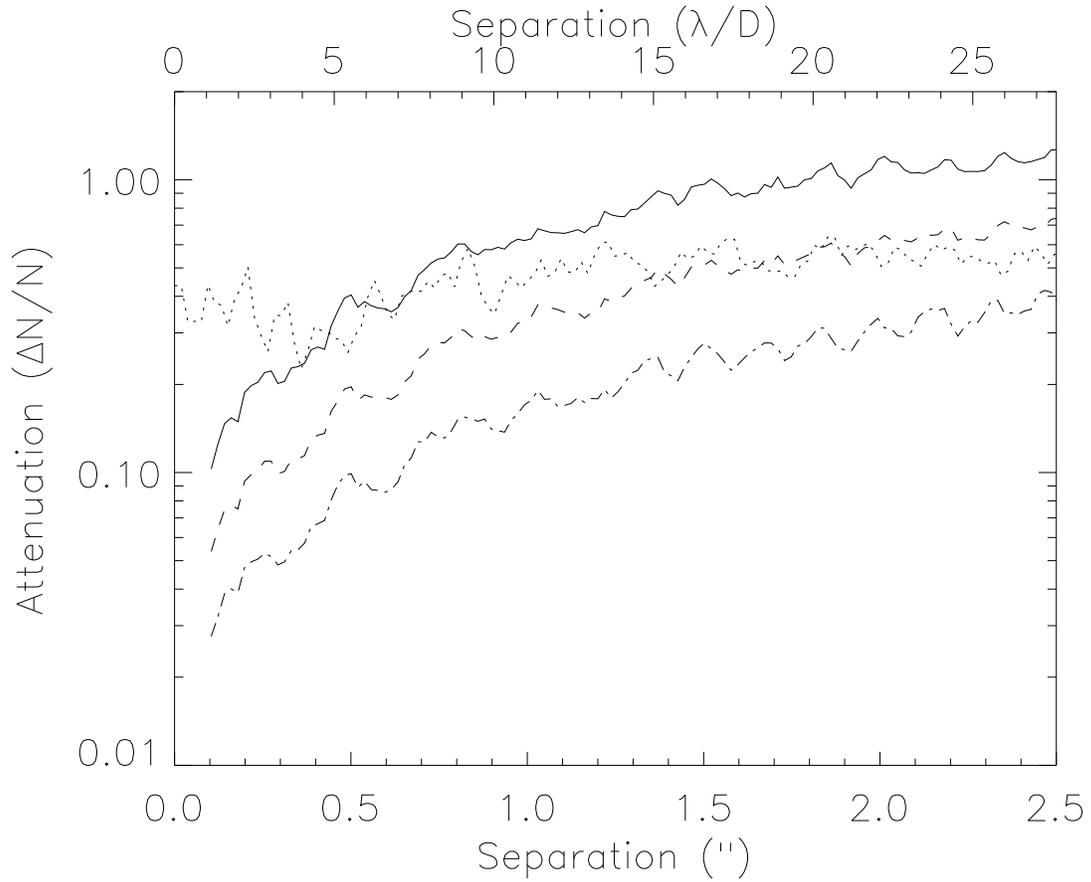}
\caption{PSF attenuation with separation for the wavefront drift effect. The solid line shows the PSF noise attenuation performance for a 1/100 wavefront drift between two PSFs, while the dashed and dot dashed lines are respectively for a 1/200 and 1/400 wavefront drift between PSFs. The dotted line shows the PSF attenuation obtained from the subtraction of two 20 minute integrations on the same 
target, separated by a 50 minute time interval.\label{pasp2003figpupshear}}
\end{figure}

A numerical simulation was performed to simulate the effect of a wavefront drift on an aberrated optical surface and to estimate the noise attenuation performance. For simplicity, the wavefront drift is assumed to be in the pupil plane, so the wavefront drift effect is simulated by shifting the phase error. A theoretical PSF is generated with the estimated 130~nm rms phase error at the pupil with $\alpha = -2.7$ at 1.580~$\mu $m. Photon noise, read noise, flat field error and atmospheric turbulence are neglected. Another PSF is then generated at the same wavelength by shifting the phase error by 1/100, 1/200 and 1/400 of the pupil diameter. The two PSFs of each set are then subtracted. Fig.~\ref{pasp2003figpupshear} shows the resulting noise attenuation. A wavefront drift tends to decorrelate higher pupil spatial frequencies more than lower spatial frequencies, reducing the PSF attenuation performance with increasing angular separation.

\begin{figure}
\plotone{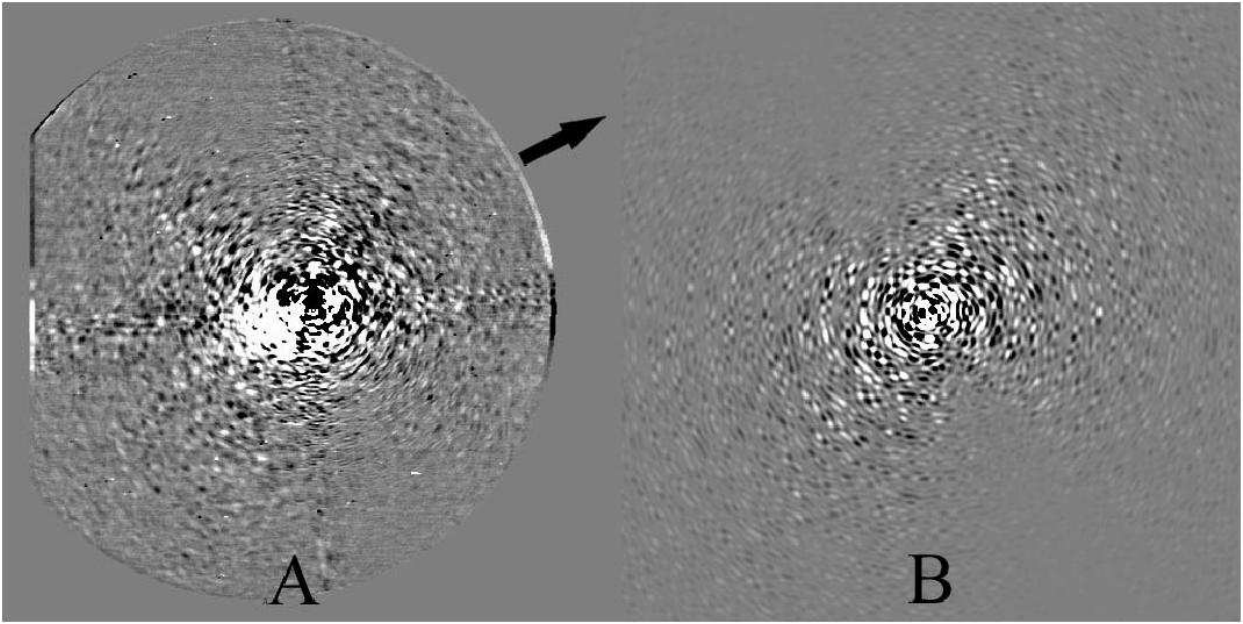}
\caption{Quasi-static PSF evolution with changing telescope pointing. A- Subtraction of two 20 minute integrations, separated by a 50 minute interval, acquired on the same object and in the same bandpass. The PSF has moved by $\sim $90~mas (in the arrow direction) between the two images mostly due to differential refraction between the AO wavefront sensor visible wavelength and TRIDENT infrared wavelengths. The two images have been registered before being subtracted. B- Subtraction of two simulated PSFs with a differential wavefront drift of 1/200 of the wavefront diameter. Each frame has a size of $9^{\prime \prime} \times 9^{\prime \prime}$ and the data are shown on an intensity scale linear within a range of $\pm3\times10^{-5}$ times the PSF peak intensity.\label{pasp2003fig8}}
\end{figure}

Evidence of the PSF drift effect is visible in TRIDENT data. For comparison, the PSF noise attenuation achieved from the subtraction of two 20 minute $\upsilon$~And images separated by a 50 minute interval between them is shown in Fig.~\ref{pasp2003figpupshear}. Fig.~\ref{pasp2003fig8} shows the subtraction of the two $\upsilon$~And images with a 1/200 wavefront drift PSF simulation for visual comparison. Note that the PSF subtraction ``structure free'' region perpendicular to the wavefront displacement in Fig.~\ref{pasp2003fig8}A is also visible in Fig.~\ref{pasp2003fig8}B. During the 50 minute time interval, the PSF moves by $\sim $90~mas, in close agreement with a numerical estimate using an atmospheric diffraction model \citep{allen1976} for 0$^{\circ}$C and 600~mm Hg. This PSF drift corresponds to a wavefront drift of $\sim $1/3000 of the wavefront diameter at the PUEO collimator and up to $\sim $1/10 of the wavefront diameter at the TRIDENT narrow band filters (last optical surface before the detector). A difference is visible at small separations between the simulated and observed noise attenuations in Fig.~\ref{pasp2003figpupshear}. This difference may be explained by the simplicity of the model which assumes that all aberrations are inside a pupil plane. This analysis shows that reference stars are of limited use for precise PSF quasi-static noise attenuation if there is no compensation for atmospheric refraction. Using an optically precise atmospheric dispersion corrector before the instrument could stabilize the wavefront path and produce a more static PSF structure that would be easier to calibrate.

\begin{figure}
\plotone{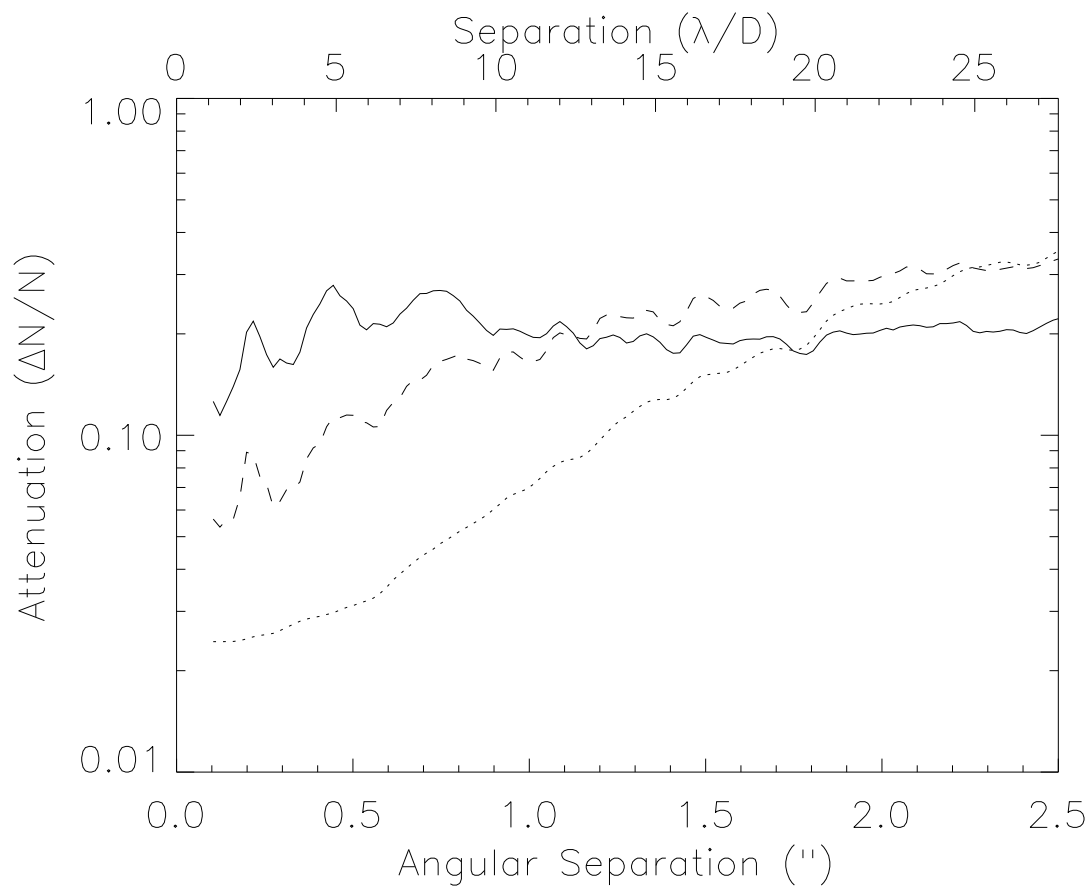}
\caption{TRIDENT PSF noise attenuation with a 1~minute calibration interval on the same target. Images are combined by odd and even number. Total integration time is 7~minutes for each combined image. The solid line shows the odd and even PSF subtraction in one wavelength. The dashed line shows the subtraction gain achieved with a reference PSF acquired simultaneously at an adjacent wavelength. The dotted line is the flat field, photon and read noises for the odd and even image subtraction with simultaneous reference PSF subtracted.\label{pasp2003fig10}}
\end{figure}

The possibility of calibrating the differential aberrations may be tested also by looking at a series of consecutive images of $\upsilon$~And acquired during the night of 2001 November 21. For this observation sequence, images were acquired every minute. For each wavelength, the seven odd numbered and the seven even numbered images were coadded separately.  The resulting images should have a good seeing correlation and the same average quasi-static aberrations. Subtraction of the two coadded images at a given wavelength should remove the average quasi-static structure and leave images dominated by atmospheric speckle noise. Further subtraction from each other of the images obtained at the different wavelengths may be expected to remove atmospheric speckle noise. Fig.~\ref{pasp2003fig10} shows the two subtraction sets compared with the estimated noise.  Each subtraction stage is seen to improve the noise attenuation significantly and yet together they are not sufficient to reach the level of the combined flatfield, photon, and read noise. This suggests that the combination of the atmosphere induced wavefront distortions and the instrument induced quasi-static wavefront distortions leads to an interference pattern that cannot be reproduced by a two stage calibration such as the one presented here.

\section{Discussion \label{discussion}}
\citet{marois2000a} derived a noise attenuation model from the Mar\'{e}chal Strehl equation \citep{marechal1947}. In light of the non-common path problem, this model needs to be modified. Following \citet{bloemhof2001,siv2002,perrin2003,bloemhof2003}, considering only phase error for simplicity, we can expand the wavefront complex exponential and write the PSF, $I$, as the following expansion

\begin{equation}
I = \sum_n a_n
\end{equation}
\begin{equation}
a_n = i^n\sum_{k=0}^{n}\frac{(-1)^{n-k}}{k!(n-k)!}(p\star^k \Phi)(p^*\star^{n-k}\Phi^*))
\end{equation}

\noindent where $p$ and $\Phi$ are respectively the Fourier transform of the pupil $P$ and phase error $\phi$. The $\star^n$ symbol is for an n-fold convolution operator, e.g., $x\star^3y = x \star y \star y \star y$. Assuming small phase errors, we conserve only the first two orders in $\Phi$ to obtain the corresponding PSFs

\begin{equation}
I \cong I_0 +2\Im\left[ p(p^*\star \Phi^*)\right] - \Re\left[ p(p^*\star^2 \Phi^*)\right] + |p\star\Phi|^2 \label{psftridentart}
\end{equation}

\noindent where $I_0$ is simply $pp^*$, the perfect diffraction figure.

The SD noise intensity attenuation can be defined as the SD over the PSF noise intensity given by $I_{\lambda_1} - I_0$

\begin{equation}
\frac{SD}{I_{\lambda_1}-I_0} \cong \frac{2\Im\left[ p(p^*\star \Delta \Phi^*)\right] + 2\Re(p\star\Phi_1)(p^*\star\Delta \Phi^*)-2\Re\left[ p(p^*\star \Phi_1^* \star \Delta \Phi^*)\right]}{2\Im\left[ p(p^*\star\Phi_1^*)\right]-\Re\left[ p(p^*\star^2\Phi^*_1)\right] + |p\star\Phi_1|^2}\label{eqtridentsd}
\end{equation}

\noindent where $\Delta \Phi$ is the Fourier transform of $\Delta \phi = \phi_2 - \phi_1$. We can simplify this equation by assuming $\Delta \Phi \ll \Phi_1$ and by considering two cases (1) near diffraction maxima ($p_{\rm{max}}$) and (2) near diffraction minima ($p_{\rm{min}} \cong 0$). For the first case, since the phase error $\Phi_1$ is small, we neglect $2\Re(p\star\Phi_1)(p^*\star\Delta \Phi^*)$, $2\Re\left[ p(p^*\star \Phi_1^* \star \Delta \Phi^*)\right]$, $\Re\left[ p(p^*\star^2\Phi^*_1)\right]$ and $|p\star\Phi_1|^2$ terms since they are quadratic in $\Phi_1$ and $\Delta \Phi$ and are thus small compared to $2\Im\left[ p(p^*\star \Delta \Phi^*)\right]$ and $2\Im\left[ p(p^*\star\Phi_1^*)\right]$ that are linear in $\Delta \Phi$ and $\Phi_1$. For the second case, we neglect terms multiplied by $p$ rather than $p\star \Phi_1$ since $p \cong 0$ near diffraction minima but not $p\star \Phi_1$. The PSF noise intensity attenuation for both diffraction maxima and minima are thus

\begin{eqnarray}
\left[ \frac{SD}{I_{\lambda_1} - I_0}\right]_{\rm{maxima}} & \cong & \frac{2\Im \left[p(p^*\star \Delta \Phi^*)\right]}{2\Im \left[ p(p^*\star \Phi_1^*)\right]} \label{eqtridentnc1}\\
\left[ \frac{SD}{I_{\lambda_1} - I_0}\right]_{\rm{minima}} & \cong & \frac{2(p^*\star \Delta \Phi^*)}{(p^*\star \Phi_1^*)}\rm{.}\label{eqtridentnc2}
\end{eqnarray}

\noindent The PSF noise attenuation is then obtained by calculating the noise rms inside annuli of increasing diameter for both the numerator and denominator of Eqs.~\ref{eqtridentnc1} and \ref{eqtridentnc2}. If we assume that $\Delta \Phi$ and $\Phi_1$ have a similar power spectrum, the noise structure of both the numerator and denominator will only differ by the ratio of the phase rms errors. We thus simply find that the PSF noise attenuations are

\begin{eqnarray}
\left[ \frac{\Delta N}{N} \right]_{\rm{maxima}} & \cong & \frac{\Delta \sigma}{\sigma} \label{eqcm01}\\
\left[ \frac{\Delta N}{N} \right]_{\rm{minima}} & \cong & 2\frac{\Delta \sigma}{\sigma}\label{eqcm02}
\end{eqnarray}

\noindent where $\Delta \sigma$ and $\sigma$ are respectively the rms noise amplitude of the non-common and common phase error. The TRIDENT experiment at CFHT has shown an attenuation of $\sim 0.5$ at 0.5$^{\prime \prime}$ (see Fig.~\ref{pasp2003graphgl526}). With the empirical fitted aberrations of 130~nm rms (assuming only phase errors), a $\Delta \sigma = 65$~nm rms is found. Each optical channel thus has $\sim 45$~nm rms ($S = 0.97$ in the $H$ band), consistent with the Strehl ratios found in the laboratory.

\begin{figure}
\plotone{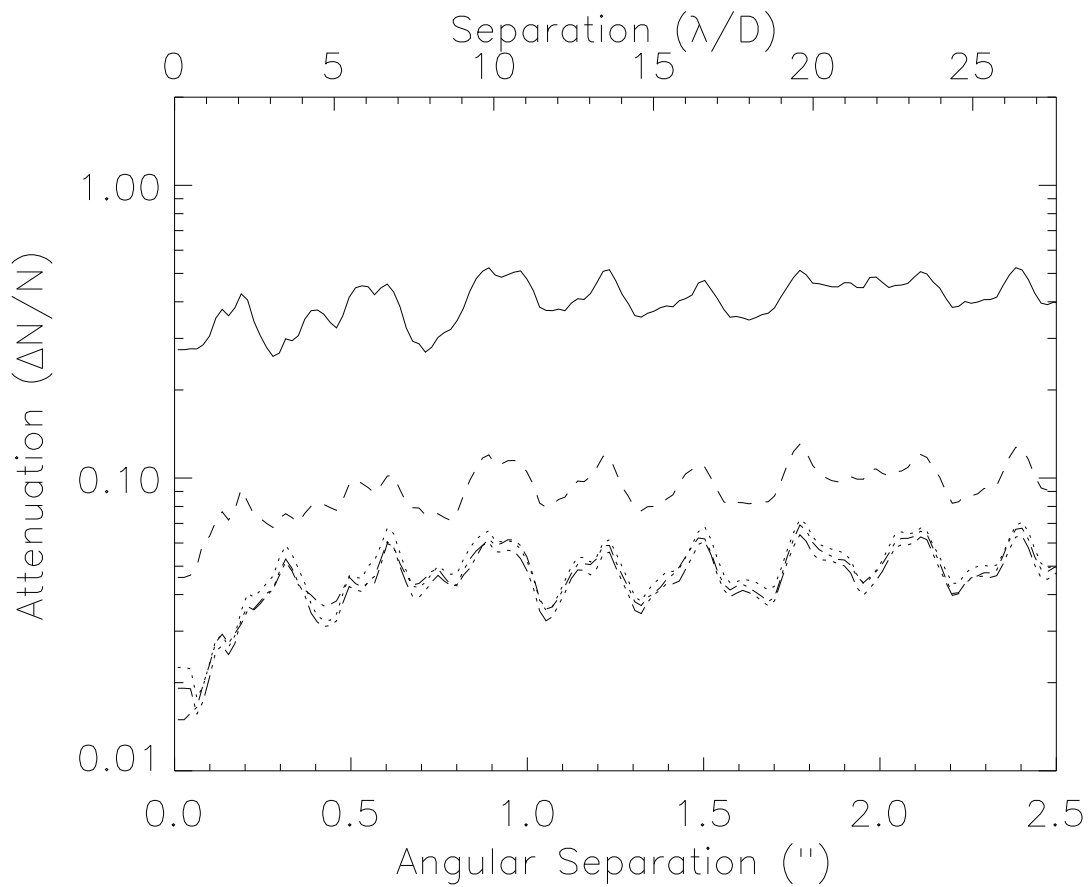}
\caption{SD PSF noise attenuation with separation for two PSFs at 1.58~$\mu $m and 1.625~$\mu $m having 130~nm rms with $\alpha = -2.7$ phase error. Dotted line shows the SD attenuation without non-common path aberration. This SD is limited by the PSF chromatic evolution. Non-common path aberrations having the same power-law distribution are gradually included in each channel: 0.2~nm rms (three dotted line), 1~nm (dot dashed line), 5~nm (dashed line) and 25~nm (solid line).\label{pasp2003fig11}}
\end{figure}

To illustrate the effect of non-common path errors, a simulation has been performed using only phase aberrations. A set of PSFs having 130~nm rms phase error with $\alpha = -2.7$ are generated at 1.58~$\mu $m and 1.625~$\mu $m. Non-common path aberrations (0.2, 1, 5 and 25~nm rms) are included using the same power-law function. Photon noise, read noise, flat field accuracy and atmospheric turbulence are neglected. Fig.~\ref{pasp2003fig11} shows the SD noise attenuation of these simulated PSFs as a function of separation. As deduced from Eq.~\ref{eqcm01}, Eq.~\ref{eqcm02} and Fig.~\ref{pasp2003fig11}, PSF noise attenuation is proportional to $\Delta \sigma/\sigma$ if non-common path aberration are larger than the PSF chromatic evolution (approximately 4~nm rms in the above simulation). The above PSF model can reproduce both laboratory and on-sky observations relatively well. It also underscores the importance of minimizing non-common path errors which imprint an uncorrelated quasi-static structure between channels that prevent precise attenuation of common path induced quasi-static speckles.

In its current implementation, TRIDENT has a detection limit competitive with other ground-based surveys done with larger telescopes.  For example, \citet{luhman2002} report a $\Delta H=10.5$ at $1^{\prime \prime}$ in 18~minutes using NICMOS and KECK while \citet{masciadri2005} achieved a $\Delta H=9.2$ at $0.5^{\prime \prime}$ in 22 minutes on the VLT AO system. \citet{close2005} report a $6\sigma$ detection sensitivity of $\Delta H=11$ at $0.5^{\prime \prime}$ in 40~minutes on the VLT, with an SSDI instrument similar to TRIDENT \citep{biller2004}. The $\sim1.5$ magnitude difference is fully accounted for by the difference in telescope size, given that companion sensitivity scales as the telescope area \citep{racine1999} in the speckle noise limited regime that applies to both CFHT and VLT observations.

As higher sensitivities are reached with SSDI, additional effects will need to be taken into account. A polychromatic speckle has a structure that results from the convolution of the monochromatic PSF with a function of the spectral distribution.  Hence, the subtraction of speckles could be limited by differences in the spectral distribution within each bandpass. These differences will come from the stellar spectrum, the Earth atmospheric transmission, the instrument optical transmission, the filter bandpass profile and the detector response with wavelength.

\begin{figure}
\plotone{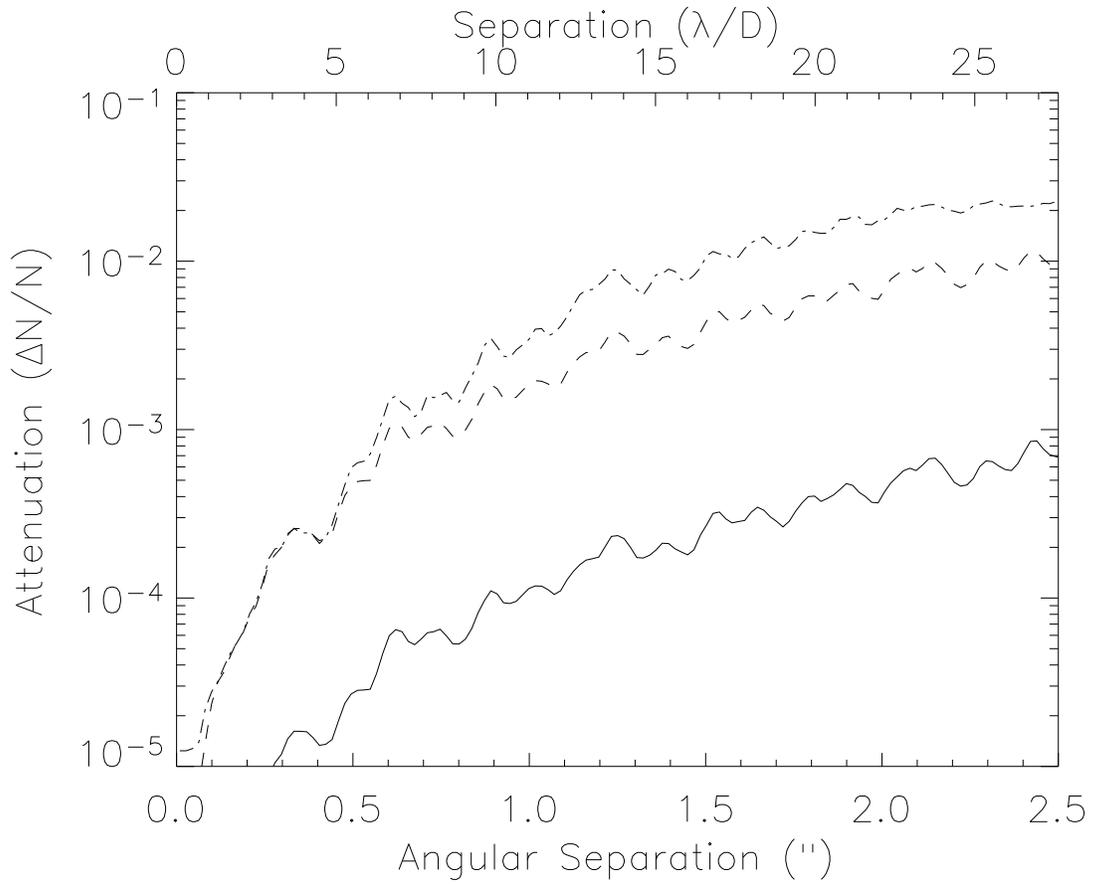}
\caption{PSF attenuation with separation for two PSFs acquired with a different spectrum. Three bandwidths are considered. The solid line shows the PSF attenuation for a 1\% bandwidth, while the dashed and dot dashed lines show, respectively, the PSF attenuation for 2\% and 5\% bandwidths.\label{pasp2003fig12}}
\end{figure}

To simulate this effect, two polychromatic PSFs are generated at 1.58~$\mu$m with 130~nm rms and $\alpha= -2.7$ with different bandwidths (1\%, 2\%, and 5\%) using the spectral distribution of the Solar spectrum at 1.58~$\mu$m and 1.625~$\mu$m (NSO/Kitt Peak FTS data produced by NSF/NOAO).  Atmospheric transmission and filter bandpass profile differences are neglected. PSFs are scaled spatially to optimize  PSF subtraction.  The PSF attenuation performance with angular separation is presented in Fig.~\ref{pasp2003fig12} for each bandwidth. As expected, the effect increases with angular separation and the bandwidth, the precise dependence being particular to the spectral distributions chosen for the simulation. Taking into account the approximations made, this simulation gives an idea of the level of noise attenuation at which a careful calibration and compensation of bandpass spectral distributions will become necessary.

The use of polarizing beam splitters in conjunction with narrowband filters, in TRIDENT, makes it possible in principle to distinguish objects that differ either in their spectral distribution, in their polarization state, or in their spatial location or extent; in the latter case because the wavelength dependent distribution of the diffracted light from the coherent wavefront of a point source differs from the wavelength independent spatial distribution of the light from incoherent sources. Given a high enough signal-to-noise ratio and an accurate correlation of diffraction patterns between optical channels, the technique implemented in TRIDENT provides enough degrees of freedom to separate the contributions from the primary star, background stars, methanated companions, and polarized emission from a circumstellar disk. Although the detection of such a disk is of high scientific interest, its presence increases the background against which a methanated companion must be detected in difference images, and it may prove desirable in practice to use a fast polarization rotator ahead of TRIDENT to ease the subtraction of polarized sources.

\section{Conclusion}
A differential imager optimized for the search, in the 1.6~$\mu$m methane absorption bandhead, for faint substellar companions, has been designed, built, and tested.  On-sky performance is limited by a quasi-static diffraction structure due to wavefront distortions originating in the adaptive optics bonnette and the imager.  This structure is attenuated by a factor of $\sim2$ when an image acquired simultaneously at an adjacent wavelength is subtracted.  Laboratory tests and simulations show that small ($\sim\lambda/20$) non-common path phase errors and amplitude modulations can explain the decorrelation of the PSF structure between bandpasses. When a reference star is used to calibrate the non-common path phase errors, a $\Delta H$ ($6\sigma$) of 9.5~magnitudes at $0.5^{\prime \prime}$ is reached on the 3.6~m CFH telescope with the PUEO adaptive optics bonnette. This calibration technique can be limited by small PSF drifts that slowly decorrelate the PSF structure if the reference PSF is not acquired with the same telescope orientation. In addition, it appears that the combination of atmosphere induced wavefront distortions and instrument induced quasi-static wavefront distortions leads to an interference pattern that is only half eliminated by the two-stage subtraction of a PSF obtained simultaneously at an adjacent wavelength and a reference PSF obtained in the same optical channel. 

A PSF noise attenuation model was presented to estimate the attenuation performance for subtraction of PSFs produced by slightly different wavefronts.  The simple equation $\Delta\sigma/\sigma$ can be used to estimate the single difference noise attenuation.

The results presented in this paper show that simultaneous differential imaging using different optical paths is very challenging, requiring the optical surface errors to be less than a few nanometers rms. A new camera design using a multi-color detector assembly consisting of microlens and micro-filter arrays has been devised \citep{marois2004b} to overcome the non-common path problem.

The authors would like to thank the referee for very useful comments and suggestions. The authors are very grateful to CFHT staff for their excellent support at the telescope. This work is supported in part through grants from the Natural Sciences and Engineering Research Council, Canada and from the Fonds Qu\'{e}b\'{e}cois de la Recherche sur la Nature et les Technologies, Qu\'{e}bec. This research was also partially performed under the auspices of the US Department of Energy by the University of California, Lawrence Livermore National Laboratory under contract W-7405-ENG-48, and also supported in part by the National Science Foundation Science and Technology Center for Adaptive Optics, managed by the University of California at Santa Cruz under cooperative agreement AST 98-76783.

\clearpage

\clearpage

\clearpage

\clearpage

\clearpage

\clearpage

\clearpage

\clearpage

\clearpage

\clearpage

\clearpage

\clearpage

\clearpage

\clearpage

\clearpage

\clearpage

\clearpage

\end{document}